\documentclass[aps,a4paper,english,nofootinbib,showpacs,preprint]{revtex4}
\usepackage{color}
\usepackage{amssymb}
\usepackage{amsmath}
\usepackage{graphicx}
\usepackage[T1]{fontenc}
\usepackage{hyperref}
\usepackage[latin1]{inputenc}
\usepackage{amsfonts}

\begin{document}

\title{Fractional fermion charges induced by vector-axial and vector gauge
potentials in planar graphene-like structures}
\author{Angel E. Obispo}
\email{ovasquez@feg.unesp.br}
\author{Marcelo Hott}
\email{marcelo.hott@pq.cnpq.br}
\affiliation{UNESP Universidade Estadual Paulista - Campus de Guaratinguetá - DFQ.\\
12516-410, Guaratinguetá-SP, Brazil.}

\begin{abstract}
We show that fermion charge fractionalization can take place in a
recently proposed chiral gauge model for graphene even in the
absence of Kekulé distortion of the graphene honeycomb lattice. We
extend the model by adding the coupling of fermions to an external
magnetic field and show that the fermion charge can be
fractionalized by means of only gauge potentials. It is shown that
the chiral fermion charge can also have fractional value. We also
relate the fractionalization of the fermion charge to the parity
anomaly in an extended Quantum Electrodynamics which involves vector
and vector-axial gauge fields.
\end{abstract}

\maketitle

\section{Introduction}

It is known that the single-particle spectrum for the charge
carriers on a honeycomb lattice described by means of a
tight-binding Hamiltonian contains two zero-energy Dirac points. The
tight-binding Hamiltonian, linearized around the Dirac points,
reveals that the kinetics of the charge carriers is governed by a
single free Dirac equation for a two-flavor spinor whose components
designate massless fermions on the two triangular sublattices of the
honeycomb lattice. A convenient coupling of the Fermi surface charge
carriers to distortions of the lattice, which preserves the
invariance of the system under time and space reversion symmetries,
opens a mass gap in the single-particle energy dispersion relation.
Recently, Hou, Chamon and Mudry \cite{HCM} have shown that when such
lattice distortions exhibits a vortex-like profile, there are zero
modes excitations in the single-particle energy spectrum. The
vorticity of the distortion, either $n\geq 1$ or $n\leq -1$ ,
determines which one of the sublattices supports $|n|$ zero modes.
From the existence of fermion zero modes and from the sublattice
symmetry, they show that the fermion quantum charge is
fractionalized, similarly to what was shown to occur in one
dimensional systems \cite{Jackiw-Rebbi}, such as polyacetylene
\cite{Schrieffer}.

Jackiw and Pi\ \cite{Jackiw-Pi} have proposed a chiral gauge theory
for graphene. They show that the system considered in \cite{HCM} is
invariant under global chiral gauge transformation and the
invariance of the system under local chiral gauge transformation
($U_{A}(1)$) is preserved if the space component of a vector-axial
gauge potential coupled in a chiral manner to the fermions is added
to the system. Such a naturally incorporated gauge potential is the
vector field which enters the phenomenological
Nielsen-Olesen-Landau-Ginzburg-Abrikosov model of a complex scalar
field minimally coupled to the $U(1)$ gauge field potential and
whose classical minimum energy solutions exhibit vortices profiles
as those binding zero-energy fermions in the scenario for fermion
charge fractionalization in \cite{HCM}. In this way, Jackiw and Pi
have specified the dynamics for the complex scalar field, which
plays the role of the distortion of the graphene lattice. Moreover,
they show that sublattice symmetry is preserved and that the fermion
zero modes, as well as the fermion charge fractionalization are
persistent even when the self-consistent vortex solution for the
gauge field is incorporated to the system.

Approximately two decades ago, Semenoff presented a physical
realization of the parity anomaly in 2+1 space-time dimensions also
in a honeycomb lattice \cite{Semenoff}. This time the lattice
contains two species of atoms and could describe for example a
monolayer of boron nitride compound. The difference in energy of
electrons on the atoms leads to a mass term for the fermions in the
effective Dirac Hamiltonian in the continuum limit. The addition of
an external gauge field coupled minimally to the fermion current
leads to an abnormal induced current for each one of the fermion
species, since the fermion mass breaks the parity symmetry. Such
abnormal current is proportional to the sign of the mass of the
fermion, and it is persistent when the mass is taken to zero. Since
the sign of the mass of one species of fermion is contrary to the
other one, the total fermion current is null, but an odd combination
of the different currents leads to a nonzero abnormal current.

For each one of the fermion species in the system considered in \cite%
{Semenoff}, the induced charge in the presence of an external
magnetic field is due to the localized fermion zero modes. The
induced charge may be fractional and proportional to the magnetic
flux on the plane \cite{Jackiw}. In the case of a vortex magnetic
field one has a finite magnetic flux and the number of
fermion zero modes is proportional to the magnetic flux \cite%
{Aharonov-Casher}.

We show that a vortex magnetic field, as in the system considered in \cite%
{Jackiw-Pi}, binds, by itself, zero-energy fermion states whose
spinor structure is equal to that found in the fermion zero modes in
\cite{HCM} and \cite{Jackiw-Pi}, that is, the zero-energy states are
supported on only one of the triangular sublattices. The breaking of
the sublattice symmetry by the fermion zero modes implies into an
induced fractional fermion charge which is proportional to the
magnetic flux. In fact, one can see from the spinor structure of
zero modes that they are eigenstates of a Dirac matrix which
anticommutes with the Dirac Hamiltonian and that, the introduction
of a staggered chemical potential term proportional to that Dirac
matrix reveals the origin of the sign of the fractional charge.
Whether the chemical potential is positive or negative implies into
the contribution of positive or negative energy states. In this way
the realization of induced fractional fermion charge in graphene is
provided in a manner very similar to that one proposed originally in
\cite{Semenoff}, that is, by means of only vector-axial gauge field.

Furthermore, by imposing the invariance of the Dirac Hamiltonian
under the usual $U(1)$ local gauge transformation, we extend the
system by minimally coupling the fermions to a vector gauge
potential. We show that a fractional fermion charge may also be
induced by a vortex-like magnetic field associated to the added
vector potential with and without the magnetic field associated to
the original vector-axial gauge potential. We analyze the discrete
symmetries of the isolated fermion zero modes and find that they are
supported on both triangular sublattices. Based on that symmetry, we
show that there is room to the fractionalization of the fermion
(chiral) charge as well.

In the next section we consider the Dirac Hamiltonian resulting from
the linearization, around the Dirac points, of the tight-binding
Hamiltonian for the fermions on the graphene honeycomb structure. In
this case we start with the extended Dirac Hamiltonian proposed in
\cite{Jackiw-Pi} and in the absence of the complex scalar field and
show the isolated fermion modes as well as the induced fermion
charge in the presence of a vortex magnetic field associated to the
vector-axial gauge potential alone. Next, we extend the system by
coupling the fermions to the usual vector gauge potential and show
the isolated fermion modes. By analyzing the discrete symmetries of
the problem we calculate the induced fermion charges in the presence
of the two kinds of magnetic vortices (or solenoids). The third
section is dedicated to a detailed analysis of the fermion states
when the magnetic fields are homogeneous and, from the behavior of
the lowest Landau level (LLL), we illustrate the origin of the
induced fermion charges. In the fifth section we obtain the
effective Chern-Simons (C-S) action in an extended Quantum
Electrodynamics (QED)in 2+1 dimensions with vector and vector-axial
gauge fields and from that we calculate the associated induced
fermion currents which recovers the fermion charges obtained in the
previous sections. The sixth section is left to the conclusions.

\section{Fermion zero modes on vortex-like magnetic fields}

The field theory Hamiltonian describing the dynamics of the
electrons on the graphene honeycomb structure in the presence of a
complex scalar field, which play the role of the lattice
distortions, together with a vector gauge potential coupled in a
chiral way to the electrons is written in a simplified version as
\begin{equation}
\mathcal{H}=\int d^{2}\vec{r}~\Psi ^{\dag }(\vec{r})\left[ -i\overset{%
\rightarrow }{\alpha }.\vec{\nabla}-\gamma _{5}\overset{\rightarrow }{\alpha
}.\overrightarrow{A_{5}}+g\beta (\varphi ^{r}-i\gamma _{5}\varphi ^{i})%
\right] \Psi (\vec{r}),  \label{H}
\end{equation}
where $\Psi (\vec{r})~$is the four component spinor
\begin{equation}
\Psi (\vec{r})=\left(
\begin{array}{c}
\psi _{+}^{b}(\vec{r}) \\
\psi _{+}^{a}(\vec{r}) \\
\psi _{-}^{a}(\vec{r}) \\
\psi _{-}^{b}(\vec{r})%
\end{array}%
\right) ,  \label{1a}
\end{equation}%
and $\overset{\rightarrow }{\alpha },~\beta $ and $\gamma _{5}\,\
$are Dirac matrices in the representation explicit below. The
superscripts in the spinor components designate the triangular
sublattice, $A$ or $B$, the electrons are supported on, while the
subscripts stand for each one of the Dirac points (the
single-particle energy spectrum obtained from the original tight
binding Hamiltonian exhibits two Dirac points). The Kekulé
distortion is represented by the complex field $\varphi (\vec{r})=|\varphi (%
\vec{r})|e^{i\chi (\vec{r})}=\varphi ^{r}(\vec{r})+i\varphi
^{i}(\vec{r})$ and $\overrightarrow{A_{5}}(\vec{r})\ $is the
vector-axial gauge potential added to make the Hamiltonian invariant
under local chiral gauge transformation $\varphi \rightarrow
e^{i2\omega (\vec{r})}\varphi $ , $\Psi \rightarrow e^{i\gamma
_{5}\omega (\vec{r})}\Psi $ \cite{Jackiw-Pi}. The matrix structure
in (\ref{H}) is revealed by means of the following gamma matrices
\begin{eqnarray}
\beta  &=&\left(
\begin{array}{cc}
0 & I \\
I & 0%
\end{array}%
\right) ,~\vec{\alpha}=\beta \overset{\rightarrow }{\gamma }=\left(
\begin{array}{cc}
\overset{\rightarrow }{\sigma } & 0 \\
0 & -\overset{\rightarrow }{\sigma }%
\end{array}%
\right) ,~  \notag \\
\alpha ^{3} &=&\beta \gamma ^{3}=\left(
\begin{array}{cc}
\sigma _{3} & 0 \\
0 & -\sigma _{3}%
\end{array}%
\right) ,~\gamma _{5}=-i\alpha ^{1}\alpha ^{2}\alpha ^{3}=\left(
\begin{array}{cc}
I & 0 \\
0 & -I%
\end{array}%
\right) ,  \label{2}
\end{eqnarray}%
where $I$ is the $2\times 2$ identity matrix and $\overset{\rightarrow }{%
\sigma }=(\sigma _{1},\sigma _{2})$ and $\sigma _{3}$ are the Pauli matrices.

Now, we consider the Hamiltonian (\ref{H}) by dropping out the
interaction term of the electrons with the Kekulé distortion and
show that if the vector-axial gauge potential has a convenient
asymptotic behavior, there will be isolated fermion zero modes in
the single-particle energy spectrum. For the sake of simplicity we
consider
\begin{equation}
A_{5}^{i}=\varepsilon ^{ij}\partial _{j}\mathcal{A}_{5}(r)\Rightarrow B_{5}=-%
\vec{\nabla}^{2}\mathcal{A}_{5}(r),  \label{3}
\end{equation}%
where $\mathcal{A}_{5}(r)$ is a scalar function of the radial coordinate
only. The time-independent Dirac equation%
\begin{equation}
\overset{\rightarrow }{\alpha }.(-i\vec{\nabla}-q\gamma _{5}\overrightarrow{%
A_{5}})\Psi (\vec{r})=E\Psi (\vec{r}),  \label{4}
\end{equation}%
can be rewritten as four equations for the spinor components (the
coupling constant $q>0$ is introduced here for future convenience).
Two of them, in a simple form, are
\begin{equation}
-2i\left[ \partial _{\overset{\_}{z}}-q\partial _{\overset{\_}{z}}\mathcal{A}%
_{5}\right] \psi _{+}^{b}=E\psi _{+}^{a}\text{ and}-2i\left[ \partial
_{z}+q\partial _{z}\mathcal{A}_{5}\right] \psi _{+}^{a}=E\psi _{+}^{b}\text{,%
}  \label{4a}
\end{equation}%
whilst the equations for $\psi _{-}^{b}$ and $\psi _{-}^{a}$ are obtained by
taking the complex conjugate of the above equations and by making the
identifications $\psi _{-}^{b}=(\psi _{+}^{b})^{\ast }$ and $\psi
_{-}^{a}=(\psi _{+}^{a})^{\ast }$. In the above equations we are using the
notation $z=x+iy$ and $\partial _{z}=(\partial _{x}-i\partial _{y})/2$. In
such a simple notation, one can easily check that the zero-energy
non-normalized eigenstates would be given by%
\begin{equation}
\psi _{+}^{b}=f(z)e^{q\mathcal{A}_{5}(r)},~\psi _{+}^{a}=h(\overset{\_}{z}%
)e^{-q\mathcal{A}_{5}(r)},~\psi _{-}^{b}=\overline{f(z)}e^{q\mathcal{A}%
_{5}(r)},\mathrm{~}\psi _{-}^{a}=\overline{h(\overset{\_}{z})}e^{-q\mathcal{A%
}_{5}(r)},  \label{5}
\end{equation}%
where $f(z)\,\ $and $h(\overset{\_}{z})$ are holomorphic functions. The
normalization of the eigenfunctions depends on the asymptotic behavior of $%
\mathcal{A}_{5}(r)$ and, from the functional dependence of the
spinor components, either $\psi _{\pm }^{b}(\vec{r})$ or $\psi _{\pm
}^{a}(\vec{r})$ is normalizable, thereby one obtains one of the
following zero-energy eigenstates
\begin{equation}
\Psi _{0}(\vec{r})=\mathcal{N}\left(
\begin{array}{c}
e^{+il\theta } \\
0 \\
0 \\
e^{-il\theta }%
\end{array}%
\right) r^{l}e^{q\mathcal{A}_{5}(r)}~~\mathrm{or}~~\Psi _{0}(\vec{r}%
)=\mathcal{N}\left(
\begin{array}{c}
0 \\
e^{-il\theta } \\
e^{+il\theta } \\
0%
\end{array}%
\right) r^{l}e^{-q\mathcal{A}_{5}(r)}\text{,}  \label{6}
\end{equation}%
where $l$ is taken to be an integer number and $\theta $ is the
angular variable in cylindrical coordinates. The degeneracy of the
zero-energy level depends on the number of values $l$ can assume and
it is linked to the magnetic flux according to the theorems proved
in \cite{Aharonov-Casher}. In this section we are concerned with
finite magnetic flux configurations only.

To show that a fractional fermion charge can be induced with such
magnetic field configurations, we note that the zero modes are
eigenfunctions of $\alpha ^{3}$ and that the Hamiltonian operator $\overset%
{\rightarrow }{\alpha }.(\vec{p}-\gamma _{5}\overrightarrow{A_{5}})$
anticommutes with the matrix $\alpha ^{3}$, then one can show that
each eigenfunction of negative energy is obtained from an
eigenfunction
of positive energy by $\Psi _{-|E|}(\vec{r})=\alpha ^{3}$ $\Psi _{|E|}\,\ $
In other words, the energy spectrum is symmetric around the
zero-energy level whose corresponding eigenfunctions are
self-conjugate. From this symmetry and from the fact that the zero
modes contain representatives of only one of the
triangular sublattices one obtains, by following the procedures in \cite%
{Jackiw}, \cite{Niemi-Semenoff} and \cite{Chamon-et-al} that the induced
fermion number is given by
\begin{equation}
N=\int d^{2}\vec{r}~\frac{1}{2}\langle 0|[\Psi ^{\dag }(\vec{r}),\Psi (\vec{r%
})]|0\rangle =\mp \frac{q\Phi _{5}}{4\pi },  \label{7}
\end{equation}%
where $\Phi _{5}=\int d^{2}\vec{r}B_{5}~$is the magnetic flux and
the upper (lower) sign on the right-hand side of (\ref{7}) comes out
when the zero modes are assigned to fermions (antifermions).

The induced fermion charge is finite provided the magnetic flux is
finite and that can be achieved by means of for example a solenoid
or vortex magnetic field, such that \cite{Jackiw} $
\mathcal{A}_{5}(r\rightarrow \infty )\sim -(\Phi _{5}/2\pi)\ln r$.
For $\Phi _{5}>0~$\ one has $\Psi _{0}^{T}(r\rightarrow \infty )\sim r^{l-%
\frac{q \Phi _{\text{\textsc{5}}}}{2\pi }}(e^{+il\theta
}~0~0~e^{-il\theta
})\,\,$and for $\Phi _{5}<0$ $\Psi _{0}^{T}(r\rightarrow \infty )\sim r^{l+%
\frac{q\Phi _{\text{\textsc{5}}}}{2\pi }}(0~e^{-il\theta }~e^{+il\theta }~0)$%
, where we are assuming that $l<[|\frac{q\Phi
_{\text{\textsc{5}}}}{2\pi }|-1] $, with $[\nu ]$ denoting the
largest integer less than $\nu $. The charge is fractional if the
magnetic flux is an integer odd number ($n_{o}$) of quantum of flux
$hv_{F}/q$, where $h$ is the Planck constant, $v_{F}$ is the Fermi
velocity and $q$ is the coupling constant associated to the chiral
coupling, that is,
\begin{equation}
Q=e\int d^{2}\vec{r}~\frac{1}{2}\langle 0|[\Psi ^{\dag }(%
\vec{r}),\Psi (\vec{r})]|0\rangle =\mp n_{o}\frac{e}{2}, \label{8a}
\end{equation}%
where $e$ is the unit of electric charge.

\subsection{Minimal coupling to a vector gauge potential, zero modes and
fermion charges fractionalization:}
The Hamiltonian (\ref{H}) is
also invariant under the usual global gauge transformation $\Psi
\rightarrow e^{i\xi }\Psi $, with $\xi $ uniform and constant. By
promoting such a transformation to a local one, a vector gauge
potential $\overrightarrow{A}$ must be added to the kinetic
term in such a way that $\overrightarrow{A}$ $\rightarrow \overrightarrow{A}+%
\overrightarrow{\nabla }\xi $.

We are now concerned with the appearance of zero-energy bound states
of the system when such a vector gauge potential is
also taken into account, namely we are going to consider the Dirac equation%
\begin{equation}
\overset{\rightarrow }{\alpha }.(-i\vec{\nabla}-q\gamma _{5}\overrightarrow{%
A_{5}}-e\overrightarrow{A})\Psi (\vec{r})=E\Psi (\vec{r}),  \label{10}
\end{equation}%
particularly when both gauge potentials engender finite magnetic
flux. Here we have also explicit the coupling constant $e>0$.

An interesting aspect of the Hamiltonian operator in (\ref{10}) is
that it anticommutes with $\alpha ^{3}$ and with $\alpha ^{3}\gamma
_{5}$. In the previous case the possible zero-energy
solutions are eingenstates of $\alpha ^{3}$, whilst in the case $%
\overrightarrow{A_{5}}=0$ and $\overrightarrow{A}$ given by%
\begin{equation}
A^{i}=\varepsilon ^{ij}\partial _{j}\mathcal{A}(r)\Rightarrow B=-\vec{\nabla}%
^{2}\mathcal{A}(r),  \label{11}
\end{equation}%
the zero-energy solutions are eingenstates of $\alpha ^{3}\gamma _{5}$. In
this case the spinor components satisfy the equations
\begin{eqnarray}
-2i\left[ \partial _{z}+e\partial _{z}\mathcal{A}\right] \psi _{+}^{a}
&=&E\psi _{+}^{b},~-2i\left[ \partial _{\overset{\_}{z}}-e\partial _{\overset%
{\_}{z}}\mathcal{A}\right] \psi _{+}^{b}=E\psi _{+}^{a},  \notag \\
2i\left[ \partial _{z}+e\partial _{z}\mathcal{A}\right] \psi _{-}^{b}
&=&E\psi _{-}^{a},~\ 2i\left[ \partial _{\overset{\_}{z}}-e\partial _{%
\overset{\_}{z}}\mathcal{A}\right] \psi _{-}^{a}=E\psi _{-}^{b},  \label{12}
\end{eqnarray}%
whose eigenfunctions for $E=0$ are given by
\begin{equation}
\Psi _{0}(\vec{r})=\mathcal{N}\left(
\begin{array}{c}
e^{+il\theta } \\
0 \\
e^{+il\theta } \\
0%
\end{array}%
\right) r^{l}e^{e\mathcal{A}(r)}~~\mathrm{or}~~\Psi
_{0}(\vec{r})=\mathcal{N}\left(
\begin{array}{c}
0 \\
e^{-il\theta } \\
0 \\
e^{-il\theta }%
\end{array}%
\right) r^{l}e^{-e\mathcal{A}(r)},\label{13}
\end{equation}%

We notice that fermion zero modes are supported on both sublattices
and are eigenstates of $\alpha ^{3}\gamma _{5}$. The energy spectrum
is symmetric around $E=0$, the negative energy eigenstates are
obtained from the positive energy ones by means of the norm
preserving operation $\Psi _{-|E|}(\vec{r})=\alpha ^{3}\gamma
_{5}\Psi _{|E|}(\vec{r})$. Thus, the induced fermion number is given
as in (\ref{7}) with $q\Phi _{5}$ replaced by $e\Phi$, with
$\Phi=\int d^{2}\vec{r}B$. By considering $\mathcal{A}%
(r\rightarrow \infty )\sim -(\Phi /2\pi)\ln r$ we have $\Psi
_{0}^{T}(r\rightarrow \infty )\sim r^{l-\frac{e\Phi }{2\pi
}}(e^{+il\theta }~0~e^{+il\theta }~0)$ for $\Phi >0$ and $\Psi
_{0}^{T}(r\rightarrow \infty
)\sim r^{l+\frac{e\Phi }{2\pi }}(0~e^{-il\theta }~0~e^{-il\theta })$ for $%
\Phi <0$. An induced fermion charge is obtained similarly to the
previous case and if the magnetic flux is quantized as an integer
odd number of quantum of flux $hv_{F}/e$ the fermion charge is
fractionalized and given by (\ref{8a}).

The fractionalization of the fermion number (charge) is also
possible in very special circumstances when both quantized magnetic
fluxes acts simultaneously. Moreover, another charge - we call it
chiral charge - can also be fractionalized. That is what we are
going to show next.

The spinor components satisfy the following equations%
\begin{eqnarray}
-2i\left[ \partial _{z}+\partial _{z}\mathcal{A}_{+}\right] \psi _{+}^{a}
&=&E\psi _{+}^{b},~-2i\left[ \partial _{\overset{\_}{z}}-\partial _{\overset{%
\_}{z}}\mathcal{A}_{+}\right] \psi _{+}^{b}=E\psi _{+}^{a},  \notag \\
2i\left[ \partial _{z}+\partial _{z}\mathcal{A}_{-}\right] \psi _{-}^{b}
&=&E\psi _{-}^{a},~2i\left[ \partial _{\overset{\_}{z}}-\partial _{\overset{%
\_}{z}}\mathcal{A}_{-}\right] \psi _{-}^{a}=E\psi _{-}^{b},  \label{14}
\end{eqnarray}%
In the equations
above we have set $\hbar =v_{F}=1$, $\mathcal{A}_{\pm }=e\mathcal{A}\pm q%
\mathcal{A}_{5}$ and, for sake of simplicity, we consider the following
asymptotic behaviors $\mathcal{A}_{5}(r\rightarrow \infty )\sim -\frac{n_{5}%
}{q}\ \ln r$ and $\mathcal{A}(r\rightarrow \infty )\sim -\frac{n}{e}\ \ln r$%
, with $n_{5}$ and $n$ positive integer numbers.

Normalized zero-energy states are obtained in some particular cases, namely
\begin{equation}
{\Psi }_{0,l_{+},l_{-}}^{n>n_{5}}(\vec{r})=\left(
\begin{array}{c}
\frac{1}{\sqrt{4\pi \int_{0}^{\infty }r^{\prime ^{2l_{\text{\textsc{+}}%
}+1}}e^{2\mathcal{A}_{\text{\textsc{+}}}(r^{\prime })}dr^{\prime }}}%
~r^{l_{+}}e^{il_{+}\theta }e^{\mathcal{A}_{+}(r)} \\
0 \\
\frac{1}{\sqrt{4\pi \int_{0}^{\infty }r^{\prime ^{2l_{\text{\textsc{-}}%
}+1}}e^{2\mathcal{A}_{\text{\textsc{-}}}(r^{\prime })}dr^{\prime }}}%
~r^{l_{-}}e^{il_{-}\theta }e^{\mathcal{A}_{-}(r)} \\
0%
\end{array}%
\right) ,~n>n_{5},\text{ }l_{\pm }=0,1,...,[n\pm n_{5}-1],  \label{15a}
\end{equation}%
or%
\begin{equation}
\Psi _{0,l_{+},l_{-}}^{n<n_{5}}(\vec{r})=\left(
\begin{array}{c}
\frac{1}{\sqrt{4\pi \int_{0}^{\infty }r^{\prime ^{2l_{\text{\textsc{+}}%
}+1}}e^{2\mathcal{A}_{\text{\textsc{+}}}(r^{\prime })}dr^{\prime }}}%
~r^{l_{+}}e^{il_{+}\theta }e^{\mathcal{A}_{+}(r)} \\
0 \\
0 \\
\frac{1}{\sqrt{4\pi \int_{0}^{\infty }r^{\prime ^{2l_{\text{\textsc{-}}%
}+1}}e^{-2\mathcal{A}_{-}(r^{\prime })}dr^{\prime }}}~r^{l_{-}}e^{-il_{-}%
\theta }e^{-\mathcal{A}_{-}(r)}%
\end{array}%
\right) ,~n<n_{5},\text{ }l_{\pm }=0,1,...,[n_{5}\pm n-1].  \label{15b}
\end{equation}%
Note that $\alpha ^{3}\gamma _{5}{\Psi }_{0,l_{+},l_{-}}^{n>n_{5}}(\vec{r})={%
\Psi }_{0,l_{+},l_{-}}^{n>n_{5}}(\vec{r})$ and $\alpha ^{3}{\Psi }%
_{0,l_{+},l_{-}}^{n<n_{5}}(\vec{r})={\Psi }_{0,l_{+},l_{-}}^{n<n_{5}}(\vec{r}%
)$. On the other hand, when the magnetic fluxes are both negative,
the zero modes are $\alpha ^{3}\gamma _{5}{\Psi
}_{0,l_{+},l_{-}}^{\left\vert n\right\vert >\left\vert
n_{5}\right\vert }(\vec{r})=-{\Psi }_{0,l_{+},l_{-}}^{\left\vert
n\right\vert >\left\vert n_{5}\right\vert }(\vec{r})$ and $\alpha ^{3}{\Psi }%
_{0,l_{+},l_{-}}^{\left\vert n\right\vert <\left\vert n_{5}\right\vert }(%
\vec{r})={\Psi }_{0,l_{+},l_{-}}^{\left\vert n\right\vert
<\left\vert n_{5}\right\vert }(\vec{r})$.

Both kinds of zero modes given by expressions (\ref{15a}) and (\ref{15b})
lead to induced fermion numbers, which depend on the magnetic fluxes as
\begin{equation}
N=\int d^{2}\vec{r}~\frac{1}{2}\langle 0|[\Psi ^{\dag }(\vec{r}),\Psi (\vec{r%
})]|0\rangle =\mp \frac{n}{2},~\mathrm{for~}\left\vert n\right\vert
>\left\vert n_{5}\right\vert ,  \label{16a}
\end{equation}%
and%
\begin{equation}
N=\int d^{2}\vec{r}~\frac{1}{2}\langle 0|[\Psi ^{\dag }(\vec{r}),\Psi (\vec{r%
})]|0\rangle =\mp \frac{n_{5}}{2},~\mathrm{for~}\left\vert n\right\vert
<\left\vert n_{5}\right\vert .  \label{16b}
\end{equation}%
Again, the upper (lower) sign on the right-hand side of (\ref{16a})
and (\ref{16b}) depends on the zero-energy states are taken as
fermions (antifermions).

We have also found that a fractional (chiral) fermion number can
also be
induced, namely%
\begin{equation}
N_{5}=\int d^{2}\vec{r}~\frac{1}{2}\langle 0|[\Psi ^{\dag
}(\vec{r}),\gamma _{5}\Psi (\vec{r})]|0\rangle =\mp
\frac{n_{5}}{2},~\mathrm{for~}\left\vert n\right\vert
>\left\vert n_{5}\right\vert ,  \label{17a}
\end{equation}%
and%
\begin{equation}
N_{5}=\int d^{2}\vec{r}~\frac{1}{2}\langle 0|[\Psi ^{\dag
}(\vec{r}),\gamma _{5}\Psi (\vec{r})]|0\rangle =\mp
\frac{n}{2},~\mathrm{for~}\left\vert n\right\vert <\left\vert
n_{5}\right\vert ,  \label{17b}
\end{equation}%
if $\left\vert n\right\vert$ and $\left\vert n_{5}\right\vert$ are
odd numbers.

This (chiral) fermion number is proportional to the non-vanishing
fractional (anomalous) fermion charge found previously in the
Boron-Nitride honeycomb structure \cite{Semenoff} where two species
of electrons which belong to the two different atoms implies into a
mass term. Such a mass term, in the representation we have been
working here, is equivalent to a staggered chemical potential. In
the next section we analyze in detail the single-particle states for
electrons in the honeycomb structure and in the presence of
homogeneous magnetic fields when a staggered chemical potential and
a parity-breaking mass term are added to the Hamiltonian.

\section{Landau levels and induced charges on uniform magnetic fields}

In this section we obtain the fermion bound states when the
electrons on the honeycomb structure are under the action of vector
gauge potentials for homogeneous magnetic fields. In this analysis
we introduce terms in the Hamiltonian density which provide gaps in
the single-particle energy spectrum. Such terms are a staggered
chemical potential $\mu \alpha ^{3}$
and a parity breaking \textquotedblleft mass\ term\textquotedblright\ $%
m_{\tau }\alpha ^{3}\gamma _{5}=m_{\tau }\beta \tau $. The staggered
chemical potential was already considered in \cite{Chamon-et-al},
\cite{Chamon-et-al2} to show the realization of irrational fermion
charge induced by scalar fields and still
maintaining the time-reversal symmetry. The parity-breaking mass matrix $%
m_{\tau }\beta \tau$ in this version of the graphene honeycomb
structure is the continuous limit of the Haldane mass (energy)
\cite{Haldane} linearized around the Dirac points \cite{tesehou}. We
consider it as constant as in \cite{Raya}, but the same term was
taken to be position-dependent (a domain wall) in \cite{Semenoff2}
to investigate how the electronic properties of graphene are
modified under such a domain wall.

When such terms are added to the Hamiltonian, the corresponding
Dirac operator anticommutes neither with $\alpha ^{3}$ nor with
$\alpha ^{3}\gamma _{5}$ and the Hamiltonian does not admit any
norm-preserving (conjugation) symmetry, that is, normalizable
negative-energy eigenstates are no longer obtained from the
normalizable positive-energy eigenstates and the lowest energy
states are not necessarily self-conjugate under the operations of
either $\alpha ^{3}$ or $\alpha ^{3}\gamma _{5}$. Nevertheless, it
may be found a non-unitary operator which conjugates negative to
positive-energy eigenstates and vice-versa, but, due to the lack of
unitarity, the density of positive-energy states is different from
the density of negative-energy states belonging to the continuous
energy spectrum and, as a consequence, an irrational fermion number
may be induced. Since we do not know exactly the fermion
single-particle energy states in a general magnetic field then we
are going to consider the case of homogeneous magnetic fields whose
single-particle energy levels are the familiar Landau levels. Once
the energy spectrum is discrete, there is no way to realize
irrational fermion number. Here the fermion number density is
proportional to the surface density of states of the LLL whose
associated eingenstates are eigenfunctions of either $\alpha ^{3}$
or $\alpha ^{3}\gamma _{5}$, depending on the magnitude of the
applied magnetic fields.

Now, the Dirac equation is%
\begin{equation}
\lbrack \overset{\rightarrow }{\alpha }.(-i\vec{\nabla}-q\gamma _{5}%
\overrightarrow{A_{5}}-e\overrightarrow{A})+\mu \alpha _{3}+m_{\tau }\beta
\tau ]\Psi (\vec{r})=E\Psi (\vec{r}),  \label{18}
\end{equation}%
where the vector potentials are written as in (\ref{3}) and (\ref{11}) with $%
\mathcal{A}(r)$ and $\mathcal{A}_{5}(r)$ given by

\begin{equation}
\mathcal{A}(r)=-\frac{Br^{2}}{4}~\mathrm{and}~\mathcal{A}_{5}(r)=-\frac{%
B_{5}r^{2}}{4}\text{ with }B,\ B_{5}>0.  \label{19}
\end{equation}%
The set of equations for the spinor components can be written as%
\begin{eqnarray}
-i\left[ 2\partial _{z}-\frac{\omega _{+}}{2}\overset{\_}{z}\right] \psi
_{+}^{a} &=&(E-m_{+})\psi _{+}^{b},~-i\left[ 2\partial _{\overset{\_}{z}}+%
\frac{\omega _{+}}{2}z\right] \psi _{+}^{b}=(E+m_{+})\psi _{+}^{a}  \notag \\
i\left[ 2\partial _{z}-\frac{\omega _{-}}{2}\overset{\_}{z}\right] \psi
_{-}^{b} &=&(E+m_{-})\psi _{-}^{a},~~i\left[ 2\partial _{\overset{\_}{z}}+%
\frac{\omega _{-}}{2}z\right] \psi _{-}^{a}=(E-m_{-})\psi _{-}^{b}
\label{20}
\end{eqnarray}%
where $\omega _{\pm }=eB\pm qB_{5}$ and $m_{\pm }=\mu \pm m_{\tau }$. This
problem is exactly solvable, but the spinorial structure of the eigenstates
of the Hamiltonian operator depends on whether $\mu >$ $m_{\tau }$ or $\mu <$
$m_{\tau }\ $as well as on whether $eB>qB_{5}$ or $eB<qB_{5}$. Hereafter, we
take $\mu $, $m_{\tau }\geq 0$ \ and $eB,\ qB_{5}>0$, although the cases
when $eB,\ qB_{5}<0$ and/or $\mu $, $m_{\tau }\leq 0$ can also be analyzed
straightforwardly. Even in the cases we are concerned here, there are four
different situations, which can be better appreciated if examined separately.

\textit{Case 1}: $qB_{5}>eB$ . The energy spectrum is $\left\vert E_{\pm
}\right\vert =\sqrt{2n\left\vert \omega _{\pm }\right\vert +m_{\pm }^{2}}$,
with $n\ \epsilon \
\mathbb{N}
$~and the eigenstates are
\begin{eqnarray}
\Psi _{+}^{E_{+}>0} &=&\frac{e^{-i\left\vert E_{+}\right\vert t}}{\sqrt{2}}%
\left(
\begin{array}{c}
\sqrt{1+\frac{m_{+}}{\left\vert E_{+}\right\vert }}R_{n,l,\omega
_{+}}(r,\theta ) \\
i\sqrt{1-\frac{m_{+}}{\left\vert E_{+}\right\vert }}R_{n-1,l+1,\omega
_{+}}(r,\theta ) \\
0 \\
0%
\end{array}%
\right) ,  \notag \\[0.12in]
\Psi _{-}^{E_{-}>0} &=&\frac{e^{-i\left\vert E_{-}\right\vert t}}{\sqrt{2}}%
\left(
\begin{array}{c}
0 \\
0 \\
-i\sqrt{1-\frac{m_{-}}{\left\vert E_{-}\right\vert }}R_{n-1,l+1,\omega _{-}}^{\ast }(r,\theta ) \\
\sqrt{1+\frac{m_{-}}{\left\vert E_{-}\right\vert }}R_{n,l,\omega_{-}}^{\ast }(r,\theta )%
\end{array}%
\right) ,  \label{21}
\end{eqnarray}

and

\begin{eqnarray}
\Psi _{+}^{E_{+}<0} &=&\frac{e^{i\left\vert E_{+}\right\vert t}}{\sqrt{2}}%
\left(
\begin{array}{c}
\sqrt{1-\frac{m_{+}}{\left\vert E_{+}\right\vert }}R_{n,l,\omega
_{+}}(r,\theta ) \\
-i\sqrt{1+\frac{m_{+}}{\left\vert E_{+}\right\vert }}R_{n-1,l+1,\omega
_{+}}(r,\theta ) \\
0 \\
0%
\end{array}%
\right)  \notag \\[0.12in]
\Psi _{-}^{E_{-}<0} &=&\frac{e^{i\left\vert E_{-}\right\vert t}}{\sqrt{2}}%
\left(
\begin{array}{c}
0 \\
0 \\
i\sqrt{1+\frac{m_{-}}{\left\vert E_{-}\right\vert }}R_{n-1,l+1,\omega _{-}}^{\ast }(r,\theta ) \\
\sqrt{1-\frac{m_{-}}{\left\vert E_{-}\right\vert }}R_{n,l,\omega_{-}}^{\ast }(r,\theta )%
\end{array}%
\right) ,  \label{22}
\end{eqnarray}%
where

\begin{equation}
R_{n,l,\omega _{\pm }}(r,\theta )=\sqrt{\left( \frac{\left\vert \omega _{\pm
}\right\vert }{2}\right) ^{l+1}\frac{n!}{\pi (n+l)!}}e^{il\theta }r^{l}e^{-%
\frac{\left\vert \omega _{\pm }\right\vert }{4}r^{2}}L_{n}^{l}(\left\vert
\omega _{\pm }\right\vert r^{2}/2)~\text{com }l\ \epsilon \
\mathbb{N}
\text{,}~  \label{23}
\end{equation}%
are the Gauss-Laguerre modes in cylindrical coordinates and
expressed in terms of the associated Laguerre polynomials
$L_{n}^{l}(\left\vert \omega _{\pm }\right\vert r^{2}/2)$.$~\,$Here
$R^{\ast }~$stands for the conjugate to $R$ whose normalization is $
\int \left\vert R_{n,l,\omega _{\pm }}(r,\theta )\right\vert
^{2}rdrd\theta =1$, and we are assuming that $R_{-1,l,\omega _{\pm
}}(r,\theta )=0$.

For $n=0$ and for the case $\mu >m_{\tau }$ $(m_{\pm }>0)$ the
lowest
energy states are%
\begin{equation}
\Psi _{+}^{E_{+}=m_{+}}=e^{-im_{+}t}\left(
\begin{array}{c}
R_{0,l,\omega _{+}}(r,\theta ) \\
0 \\
0 \\
0%
\end{array}%
\right) ,~~\Psi _{-}^{E_{-}=m_{-}}=e^{-im_{-}t}\left(
\begin{array}{c}
0 \\
0 \\
0 \\
R_{0,l,\omega _{-}}^{\ast }(r,\theta )%
\end{array}%
\right) .  \label{25}
\end{equation}%
We notice that when $m_{\tau }\rightarrow 0$, the energy states
above are degenerate such that one has only one eigenspinor with the
same structure found in (\ref{15b}), which is normalizable even when
$\mu
\rightarrow 0$, that is%
\begin{equation}
\Psi ^{E_{0}=\mu }=\left(
\begin{array}{c}
\psi _{+}^{b}(\vec{r},t) \\
0 \\
0 \\
\psi _{-}^{b}(\vec{r},t)%
\end{array}%
\right) =\frac{e^{-i\mu t}}{\sqrt{2}}\left(
\begin{array}{c}
R_{0,l,\omega _{+}}(r,\theta ) \\
0 \\
0 \\
R_{0,l,\omega _{-}}^{\ast }(r,\theta )%
\end{array}%
\right) ,  \label{25a}
\end{equation}

Had we started with $m_{\tau }=0$ and $\mu <0$ the LLL would have
energy $E_{0}=\mu <0$ and the corresponding eigenspinor would be
\begin{equation}
\Psi ^{E_{0}=-\left\vert \mu \right\vert }=\left(
\begin{array}{c}
\psi _{+}^{b}(\vec{r},t) \\
0 \\
0 \\
\psi _{-}^{b}(\vec{r},t)%
\end{array}%
\right) =\frac{e^{i\left\vert \mu \right\vert t}}{\sqrt{2}}\left(
\begin{array}{c}
R_{0,l,\omega _{+}}(r,\theta ) \\
0 \\
0 \\
R_{0,l,\omega _{-}}^{\ast }(r,\theta )%
\end{array}%
\right).  \label{26}
\end{equation}%
Then the induced fermion density numbers, for $\mu \rightarrow 0$,
are given by:
\begin{equation}
\rho =\frac{1}{2}\left\langle 0|\left[ \psi ^{\dagger },\psi \right]
|0\right\rangle =-\mathrm{sgn(}\mu \mathrm{)}\frac{qB_{5}}{4\pi },
\label{27.a}
\end{equation}%
and
\begin{equation}
\rho _{5}=\frac{1}{2}\left\langle 0|\left[ \psi ^{\dagger },\gamma _{5}\psi %
\right] |0\right\rangle =-\mathrm{sgn(}\mu \mathrm{)}\frac{eB}{4\pi }\text{.}
\label{27.b}
\end{equation}%

The above results are compatible with those found in the previous
section, (\ref{16b}) and (\ref{17b}), which reveals that the fermion
numbers are proportional to the flux of the magnetic field. The
dependence on the $\mathrm{sgn(}\mu \mathrm{)}$ is explained by the
fact that when $\mu $ is taken initially as positive, the LLL is
occupied by electrons on the conduction band ($E=\mu >0$), whereas
when $\mu <0$ the LLL is occupied by electrons on the valence band.
The above result is only valid for $\mu \rightarrow 0$, otherwise
there would be a contribution (parity invariant contribution) coming
from the higher Landau levels.

Although there is no norm-preserving operator which conjugates
negative to positive-energy eigenstates, when $\mu \neq 0$, the
energy spectrum is discrete and the eigenstates can be normalized,
such that the density of states from the positive and negative
energy spectrum (except for the isolated mode $E=\mu $) is equal to
each other. Then their contribution vanishes when one calculates the
fermion numbers themselves and it is possible to show that the
fermion numbers are proportional to the magnetic flux. In fact, the
results in (\ref{27.a}) and (\ref{27.b}) express that the fermion
density numbers are equal to the surface density of states of each
Landau level (or the surface density of zero modes) and, as a
consequence, the degeneracy is proportional to the magnetic flux.

We emphasize that in this case, as was already revealed in equations (\ref%
{16b}) and (\ref{17b}), the fermion charge is proportional to the
flux of the (axial) magnetic field, whilst the fermion chiral charge
is proportional to the flux of the magnetic field. This will be
explored in the next section to analyse the effective action for the
gauge fields in the relativistic context, where we will show that
the derivative expansion approximation to the effective action leads
to a crossed Chern-Simons term.

The next case we treat is well known and has been recently analyzed in \cite%
{Raya} by resorting to another representation for the reducible
gamma matrices in 2+1 (space-time) dimensions. Although they do not
consider the vector-axial gauge potential, their results are more
general than ours in the sense that they take inhomogeneous magnetic
fields and analyze also the behavior of the chiral condensates.

\textit{Case 2}: $qB_{5}<eB$ . The energy spectrum is identical to that in
the previous case and the eingestates are expressed as
\begin{eqnarray}
\Psi _{+}^{E_{+}>0} &=&\frac{e^{-i\left\vert E_{+}\right\vert t}}{\sqrt{2}}%
\left(
\begin{array}{c}
\sqrt{1+\frac{m_{+}}{\left\vert E_{+}\right\vert }}R_{n,l,\omega
_{+}}(r,\theta ) \\
i\sqrt{1-\frac{m_{+}}{\left\vert E_{+}\right\vert }}R_{n-1,l+1,\omega
_{+}}(r,\theta ) \\
0 \\
0%
\end{array}%
\right)  \notag \\
\Psi _{-}^{E_{-}>0} &=&\frac{e^{-i\left\vert E_{-}\right\vert t}}{\sqrt{2}}%
\left(
\begin{array}{c}
0 \\
0 \\
\sqrt{1-\frac{m_{-}}{\left\vert E_{-}\right\vert }}R_{n,l,\omega
_{-}}(r,\theta ) \\
-i\sqrt{1+\frac{m_{-}}{\left\vert E_{-}\right\vert }}R_{n-1,l+1,\omega
_{-}}(r,\theta )%
\end{array}%
\right)  \label{28}
\end{eqnarray}%
\begin{eqnarray}
\Psi _{+}^{E_{+}<0} &=&\frac{e^{i\left\vert E_{+}\right\vert t}}{\sqrt{2}}%
\left(
\begin{array}{c}
\sqrt{1-\frac{m_{+}}{\left\vert E_{+}\right\vert }}R_{n,l,\omega
_{+}}(r,\theta ) \\
-i\sqrt{1+\frac{m_{+}}{\left\vert E_{+}\right\vert }}R_{n-1,l+1,\omega
_{+}}(r,\theta ) \\
0 \\
0%
\end{array}%
\right)  \notag \\
\Psi _{-}^{E_{-}<0} &=&\frac{e^{i\left\vert E_{-}\right\vert t}}{\sqrt{2}}%
\left(
\begin{array}{c}
0 \\
0 \\
\sqrt{1+\frac{m_{-}}{\left\vert E_{-}\right\vert }}R_{n,l,\omega
_{-}}(r,\theta ) \\
i\sqrt{1-\frac{m_{-}}{\left\vert E_{-}\right\vert }}R_{n-1,l+1,\omega
_{-}}(r,\theta )%
\end{array}%
\right)  \label{29}
\end{eqnarray}

For $\mu =0$ and $m_{\tau }>0$ the eigenstates of the LLL is
\begin{equation}
\Psi ^{E=m_{\tau }}=\left(
\begin{array}{c}
\psi _{+}^{b}(\vec{r},t) \\
0 \\
\psi _{-}^{a}(\vec{r},t) \\
0%
\end{array}%
\right) =\frac{e^{-im_{\tau }t}}{\sqrt{2}}\left(
\begin{array}{c}
R_{0,l,\omega _{+}}(r,\theta ) \\
0 \\
R_{0,l,\omega _{-}}(r,\theta ) \\
0%
\end{array}%
\right) ~~,  \label{30}
\end{equation}%
and exhibits the same spinor structure as the one in (\ref{15a}), whilst for $%
\mu =0$ and $m_{\tau }<0$ we have
\begin{equation}
\Psi ^{E=-\left\vert m_{\tau }\right\vert }=\left(
\begin{array}{c}
\psi _{+}^{b}(\vec{r},t) \\
0 \\
\psi _{-}^{a}(\vec{r},t) \\
0%
\end{array}%
\right) =\frac{e^{i\left\vert m_{\tau }\right\vert t}}{\sqrt{2}}\left(
\begin{array}{c}
R_{0,l,\omega _{+}}(r,\theta ) \\
0 \\
R_{0,l,\omega _{-}}(r,\theta ) \\
0%
\end{array}%
\right) ~~.  \label{30b}
\end{equation}

In obtaining the induced fermion charges it is convenient to
separate the contribution of the LLL from the higher LL, which also
contribute in this case as well, but the last contributions vanishes when $%
m_{\tau }\rightarrow 0$ and one finds:
\begin{eqnarray}
\rho  &=&\frac{1}{2}\left\langle 0|\left[ \psi ^{\dagger },\psi \right]
|0\right\rangle =-\mathrm{sgn(}m_{\tau }\mathrm{)}\frac{eB}{4\pi }\text{,}
\label{a.2} \\
\rho _{5} &=&\frac{1}{2}\left\langle 0|\left[ \psi ^{\dagger },\gamma
_{5}\psi \right] |0\right\rangle =-\mathrm{sgn(}m_{\tau }\mathrm{)}\frac{%
qB_{5}}{4\pi }\text{,}  \label{a.3}
\end{eqnarray}%
which are consistent with (\ref{16a}) and (\ref{17a}).

\section{Induced fermion charges from field theoretical calculations}
As one knows QED in 2+1 space-time dimensions can be enriched by
adding a topological mass to the gauge field \cite{Jackiw2} and that
such a C-S term can be generated by means of perturbation
calculation \cite{Redlich}. The C-S term comes from the first-order
in external momentum contribution of the vacuum polarization
diagram.

Here, we show that C-S terms can also be induced in the extended
QED, whose Lagrangian density is%
\begin{equation}
\mathcal{L}=\bar{\Psi}~[\gamma ^{\mu }(\partial _{\mu }+eA_{\mu }+gA_{\mu
}^{(5)})-\mu \gamma ^{3}-m_{\tau }\tau ]\Psi .  \label{31}
\end{equation}%
We also compute the fermion induced currents from the C-S effective
action.

\begin{figure}[h]
\includegraphics[width=16.0cm]{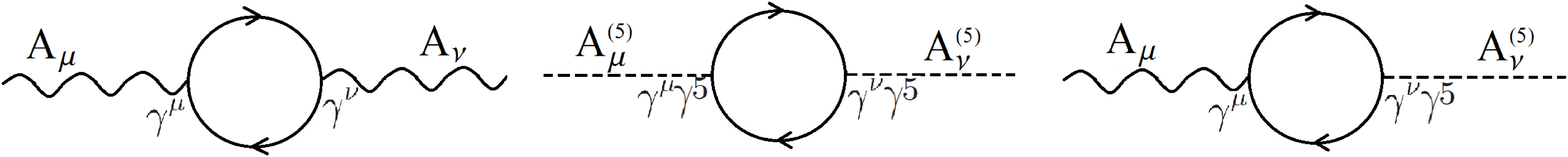}
\caption{}\label{diagrams}
\end{figure}

The vacuum polarization diagrams we have to consider are those shown
in FIG. \ref{diagrams} which correspond to the vacuum polarization
operator
\begin{equation}
\Pi ^{\mu \nu }(k)=-i\int \frac{d^{3}p}{(2\pi )^{3}}{\textrm{tr}}\left[ \frac{1}{(\slash\!\!\!%
{p}+\slash\!\!\!{k})-\mu \gamma ^{3}-m_{\tau }\tau }\gamma ^{\mu }(e+q\gamma ^{5})%
\frac{1}{\slash\!\!\!{p}-\mu \gamma ^{3}-m_{\tau }\tau }\gamma ^{\nu
}(e+q\gamma ^{5})\right] ,  \label{31b}
\end{equation}
where $S_{F}(p)=(\slash\!\!\!{p}-\mu \gamma _{3}-m_{\tau }\tau
)^{-1}$ is the fermion propagator given in terms of momenta and
under a staggered chemical potential and parity-breaking mass.

Here, as in the previous section, we are going to consider the
contributions of the staggered chemical potential and of the
parity-breaking mass separately. This is not only because we want to
see the dependence of the fermion currents on the signs of $\mu
~$and $m_{\tau }$ as in the previous section, but also because the
fermion propagator can be easily handled when the approximations
\begin{equation}
S_{F}(p)\approx S_{F,\mu
}(p)=\frac{1}{\displaystyle{\slash\!\!\!{}{p}-\mu \gamma ^{3}}}~(\mu
>>m_{\tau })  \label{34}
\end{equation}%
and%
\begin{equation}
S_{F}(p)\approx S_{F,m_{\tau
}}(p)=\frac{1}{\displaystyle{\slash\!\!\!{}{p}-m_{\tau }\tau
}}\text{ }(\mu <<m_{\tau })  \label{35}
\end{equation}%
are employed.

In both cases we look for the term which is first-order in the
external momentum, namely $\Pi _{1}^{\mu \nu }\sim \varepsilon ^{\mu
\nu \alpha }k_{\alpha }$. By taking the first approximation for the
fermion propagator, Eq. (\ref{34}), we note that the third diagram
in FIG. \ref{diagrams} is the only one which gives a contribution to
the C-S terms. This can
be appreciated through the following expression for $\Pi ^{\mu \nu }$%
\begin{equation}
\Pi ^{\mu \nu }(k)=-i\int \frac{d^{3}p}{(2\pi )^{3}}{\textrm{tr}} \left[ \frac{(\slash\!\!\!{p}+%
\slash\!\!\!{k})+\mu \gamma ^{3}}{(p+k)^{2}-\mu ^{2}}\gamma ^{\mu }(e+q\gamma ^{5})%
\frac{\slash\!\!\!{p}+\mu \gamma ^{3}}{p^{2}-\mu ^{2}}\gamma ^{\nu }(e+q\gamma ^{5})%
\right] ,  \label{35b}
\end{equation}
and by noting that the Levi-Civita tensor comes from $\rm{tr}
(\gamma ^{\mu }\gamma ^{\nu }\gamma ^{\alpha }\tau )=-4i\varepsilon
^{\mu \nu \alpha }$.
By isolating the relevant terms to the C-S contribution one finds%
\begin{eqnarray}
\Pi _{1}^{\mu \nu } &=&-\frac{\mu eq}{2\pi ^{3}}\varepsilon ^{\mu
\nu \alpha
}k_{\alpha }\lim_{k\rightarrow 0}\int d^{3}p\frac{1}{(p^{2}-\mu ^{2})}\frac{%
1}{(p+k)^{2}-\mu ^{2}}=  \notag \\
&=&-\frac{\mu eq}{2\pi ^{3}}\varepsilon ^{\mu \nu \alpha }k_{\alpha
}\lim_{k\rightarrow 0}\int_{0}^{1}dz\int d^{3}p\frac{1}{[p^{2}-\mu
^{2}+k^{2}z(1-z)]^{2}}=  \notag \\
&=&i\frac{\mu eq}{2\pi }\varepsilon ^{\mu \nu \alpha }k_{\alpha
}\lim_{k\rightarrow 0}\int_{0}^{1}\frac{dz}{[-k^{2}z^{2}+k^{2}z-\mu
^{2}]^{1/2}}=i\frac{eq}{2\pi }\varepsilon ^{\mu \nu \alpha }k_{\alpha }\frac{%
\mu }{\left\vert \mu \right\vert }.  \label{37}
\end{eqnarray}

From the effective action up to second-order on the gauge fields
corresponding to the third diagram%
\begin{equation}
S_{eff}^{(2)}=\frac{1}{2}\int d^{3}x\left\{ A_{\mu }(x)\Pi ^{\mu \nu
}(k_{\alpha }=i\partial _{\alpha })A_{\nu }^{(5)}(x)+A_{\mu }^{(5)}(x)\Pi
^{\mu \nu }(k_{\alpha }=i\partial _{\alpha })A_{\nu }(x)\right\} ,
\label{38}
\end{equation}%
one obtains a crossed C-S term, namely%
\begin{equation}
S_{eff}^{(2,1)}=-\frac{eq}{4\pi }\frac{\mu }{\left\vert \mu \right\vert }%
~\varepsilon ^{\mu \nu \alpha }\int d^{3}x\left\{ A_{\mu }(x)\partial
_{\alpha }A_{\nu }^{(5)}(x)+A_{\mu }^{(5)}(x)\partial _{\alpha }A_{\nu
}(x)\right\} .  \label{39}
\end{equation}

The induced fermion current density is obtained from this
contribution to the effective action through%
\begin{eqnarray}
\left\langle j^{\mu }\right\rangle &=&\frac{\delta S_{ef}^{(2,1)}}{\delta
A_{\mu }}=-\frac{eq}{4\pi }\frac{\mu }{\left\vert \mu \right\vert }%
\varepsilon ^{\mu \nu \alpha }\partial _{\alpha }A_{\nu }^{(5)}  \notag \\
&=&-\frac{eq}{4\pi }\frac{\mu }{\left\vert \mu \right\vert }F_{(5)}^{\mu },
\label{40}
\end{eqnarray}%
where $F_{(5)}^{\mu }=(-1/2)\varepsilon ^{\mu \nu \alpha }F_{\nu
\alpha (5)}$ is the dual of the (axial) field strength, whereas the
(axial) fermion current density can be obtained as
\begin{eqnarray}
\left\langle j_{5}^{\mu }\right\rangle &=&\frac{\delta S_{ef}^{(2,1)}}{%
\delta A_{\mu }^{(5)}}=-\frac{eq}{4\pi }\frac{\mu }{\left\vert \mu
\right\vert }\varepsilon ^{\mu \nu \alpha }\partial _{\alpha }A_{\nu }
\notag \\
&=&-\frac{eq}{4\pi }\frac{\mu }{\left\vert \mu \right\vert }F^{\mu }.
\label{41}
\end{eqnarray}
By taking the time-component of (\ref{40}) and (\ref{41}) one
recovers the
results for the fermion charge densities, equations (\ref{27.a}) and (\ref%
{27.b}), respectively.

Now, we consider the effective C-S action coming from the
contribution of the first two diagrams. We notice that such
contributions
are obtained by taking the second approximate fermion propagator given in (%
\ref{35}). Here we calculate both contributions together, that is%
\begin{equation}
\tilde{\Pi}^{\mu \nu }(k)=-i\int \frac{d^{3}p}{(2\pi )^{3}} {\rm{tr}} \left[ \frac{(%
\slash\!\!\!{p}+\slash\!\!\!{k})+m_{\tau }\tau }{(p+k)^{2}-m_{\tau
}^{2}}\gamma ^{\mu }(e+q\gamma ^{5})\frac{\slash\!\!\!{p}+m_{\tau
}\tau }{p^{2}-m_{\tau }^{2}}\gamma ^{\nu }(e+q\gamma ^{5})\right] .
\label{41b}
\end{equation}%
This time there will be no crossed C-S terms because $\rm{tr}
(\gamma ^{\mu }\gamma ^{\nu }\gamma ^{\alpha }\tau )=-4i\varepsilon
^{\mu \nu \alpha
}$ is obtained from the contributions proportional to $m_{\tau }$ and to $%
e^{2}$ and $q^{2}$, thus one has

\begin{equation}
\tilde{\Pi}_{1,1st}^{\mu \nu }=i\frac{m_{\tau }}{\left\vert m_{\tau
}\right\vert }\frac{e^{2}}{2\pi }\varepsilon ^{\mu \nu \alpha }k_{\alpha }%
\text{ and }\tilde{\Pi}_{1,2nd}^{\mu \nu }=i\frac{m_{\tau }}{\left\vert
m_{\tau }\right\vert }\frac{q^{2}}{2\pi }\varepsilon ^{\mu \nu \alpha
}k_{\alpha },  \label{43}
\end{equation}%
for the first and second diagram, respectively, and the corresponding C-S
effective action is%
\begin{equation}
\tilde{S}_{eff}^{(2,1)}=-\frac{1}{4\pi }\frac{m_{\tau }}{\left\vert m_{\tau
}\right\vert }~\varepsilon ^{\mu \nu \alpha }\int d^{3}x~\left\{ e^{2}A_{\mu
}(x)\partial _{\alpha }A_{\nu }(x)+q^{2}A_{\mu }^{(5)}(x)\partial _{\alpha
}A_{\nu }^{(5)}(x)\right\} .  \label{44}
\end{equation}%
From (\ref{44}) one obtains the induced fermion currents
\begin{equation*}
\left\langle \tilde{j}^{\mu }\right\rangle =\frac{\delta \tilde{S}%
_{ef}^{(2,1)}}{\delta A_{\mu }}=-\frac{e^{2}}{4\pi }\frac{m_{\tau }}{%
\left\vert m_{\tau }\right\vert }F^{\mu }\text{ and}~\left\langle j_{5}^{\mu
}\right\rangle =\frac{\delta S_{ef}^{(2,1)}}{\delta A_{\mu }^{(5)}}=-\frac{%
q^{2}}{4\pi }\frac{m_{\tau }}{\left\vert m_{\tau }\right\vert }F_{(5)}^{\mu
},
\end{equation*}%
whose time-components recovers the results (\ref{a.2}) and (\ref{a.3})
respectively.

\section{Conclusions}
We have obtained induced fermion charges in a continuum chiral
theory for massless planar electrons in a honeycomb structure by
means of only vector gauge potentials. For some configurations of
the vector gauge potentials, particularly quantized magnetic
vortices, we show that the induced charges can have fractional
values. Since isolated and localized zero-energy states together
with symmetry of sublattices are crucial for the appearance of
fractional charges, we have analyzed the spinorial structure of the
zero-modes as well as the spectral-symmetry conjugation of the
first-quantized Hamiltonian. Moreover, in order to understand and
highlight the origin of zero-energy states, that is, if they come
from electrons on either valence or conduction bands we have studied
in detail the energy eigenstates of fermions in homogeneous magnetic
fields by taking also into account a staggered chemical potential
($\mu$) and a parity-breaking mass term ($m_{\tau}$), which open a
mass gap in the energy spectrum (Landau levels). By taking
$\mu$,$m_{\tau}\rightarrow 0$, we calculate the induced fermion
charges and the results are in consonance with the results found
when the electrons are under the influence of finite magnetic
fluxes.

In pursuit of the realization of quantum anomalies in condensed
matter systems, we also also discuss, through field-theoretical
calculation, the relation of induced fermion currents with the
parity-anomaly in an extended Quantum Electrodynamics which involves
a vector and a vector-axial gauge fields. Unfortunately, this
relation is not established beyond doubt, since the parity-anomaly
is only realized if parity symmetry-breaking terms, such as the
staggered chemical potential and the Haldane mass, are present in
the physical system, and that is not the case in graphene.

\section{Acknowledgments}
AEO thanks to Brazilian funding agency CAPES for the scholarship
under PEC-PG program. This work is also partially supported by CNPq
(procs. 482043/2011-3, 304352/2009-8).

\end{document}